\documentclass[twocolumn,showpacs,preprintnumbers,amsmath,amssymb]{revtex4}
\usepackage{graphicx}% Include figure files
\usepackage{dcolumn}% Align table columns on decimal point
\usepackage{bm}% bold math

\begin{document}

\renewcommand{\vec}[1]{\ensuremath{\boldsymbol{#1}}}

\newcommand{\Ca}{$^{43}$Ca$^+$ }

\newcommand{\ofs}{\omega_{\text{fs}}}
\newcommand{\1}{S\ensuremath{_{1/2}}}
\newcommand{\2}{P\ensuremath{_{1/2}}}
\newcommand{\4}{P\ensuremath{_{3/2}}}
\newcommand{\3}{D\ensuremath{_{3/2}}}
\newcommand{\5}{D\ensuremath{_{5/2}}}
\newcommand{\trans}{\ensuremath{\leftrightarrow}}

%\preprint{APS/123-QED}

\title{Measurement of the hyperfine structure of the $\mathbf{S_{1/2}-D_{5/2}}$ transition in \Ca}% Force line breaks with \\

\author{J. Benhelm$^1$}
 \email{jan.benhelm@uibk.ac.at}
\author{G. Kirchmair$^1$}
\author{U. Rapol$^1$}
\altaffiliation{Present address: GE India Technology Center,
Bangalore, India}
\author{T. K\"orber$^1$}
\author{C. F. Roos$^{1,2}$}
\author{R. Blatt$^{1,2}$}
 \homepage{http://www.quantumoptics.at}
\affiliation{$^1$Institut f\"ur Experimentalphysik, Universit\"at Innsbruck, Technikerstr.~25, A-6020 Innsbruck, Austria\\
$^2$Institut f\"ur Quantenoptik und Quanteninformation,
\"Osterreichische Akademie der Wissenschaften, Otto-Hittmair-Platz 1, A-6020 Innsbruck, Austria}%

\date{\today}% It is always \today, today,

\begin{abstract}
The hyperfine structure of the $4s\,^2\!S_{1/2}-3d\,^2\!D_{5/2}$
quadrupole transition at 729 nm in $^{43}$Ca$^+$ has been
investigated by laser spectroscopy using a single trapped
$^{43}$Ca$^+$ ion. We determine the hyperfine structure constants
of the metastable level as $A_{D_{5/2}}=-3.8931(2)\,$MHz and
$B_{D_{5/2}}=-4.241(4)\,$ MHz. The isotope shift of the transition
with respect to $^{40}$Ca$^+$ was measured to be
$\Delta_{iso}^{43,40}=4134.713(5)$ MHz. We demonstrate
% that $^{43}$Ca$^+$ has transitions
the existence of transitions that become independent of the
first-order Zeeman shift at non-zero low magnetic fields. These
transitions might be better suited for building a frequency
standard than the well-known 'clock transitions' between m=0
levels at zero magnetic field.
\end{abstract}

% PACS numbers
                 % 32.80.Pj: Opt.cooling of atoms;trapping
                 % 42.62.Fi: Laser spectroscopy
                 % 31.30.Gs: Hyperfine interactions and isotope shifts
                 % 32.60.+i: Zeeman and Stark effects

%\keywords{hyperfine structure, isotope shift, clock states}%Use showkeys class option if keyword
\maketitle

\section{\label{sec:level1}Introduction}

In recent years, optical frequency standards based on single
trapped ions and neutral atoms held in optical lattices have made
remarkable progress \cite{Oskay06,Boyd06} towards achieving the
elusive goal \cite{Dehmelt82} of a fractional frequency stability
of $10^{-18}$. In $^{199}$Hg$^+$, $^{27}$Al$^+$, $^{171}$Yb$^+$,
$^{115}$In$^+$, and $^{88}$Sr$^+$, optical frequencies of
dipole-forbidden transitions have been
measured\cite{Oskay06,Schmidt06,Schneider05,vonZanthier00,Margolis04}.
Among the singly-charged alkali-earth ions, the odd isotope
$^{43}$Ca$^+$ has been discussed as a possible optical frequency
standard \cite{Champenois04,Hosokawa05} because of its nuclear
spin $I=7/2$ giving rise to transitions
$4s\,^2\!S_{1/2}(F,m_F=0)\leftrightarrow
3d\,^2\!D_{5/2}(F^\prime,m_{F^\prime}=0)$ that are independent of
the first-order Zeeman effect. While the hyperfine splitting of
the $S_{1/2}$ ground state has been precisely measured
\cite{Werth94}, the hyperfine splitting of the metastable
$D_{5/2}$ has been determined with a precision of only a few MHz
so far \cite{Noertershaeuser98}. A precise knowledge of the
$S_{1/2}\leftrightarrow D_{5/2}$ transition is also of importance
for quantum information processing based on $^{43}$Ca$^+$
 \cite{Steane97}. In these experiments where quantum information is
encoded in hyperfine ground states, the quadrupole transition can
be used for initialization of the quantum processor and for
quantum state detection by electron shelving.

This paper describes the measurement of the hyperfine constants of
the $D_{5/2}$ level by probing the quadrupole transition of a
single trapped ion with a narrow-band laser. Our results confirm
previous measurements and reduce the error bars on $A_{D_{5/2}}$
and $B_{D_{5/2}}$ by more than three orders of magnitude. In
addition, we precisely measure the isotope shift of the transition
with respect to $^{40}$Ca$^+$.

With a precise knowledge of the hyperfine structure constants at
hand, the magnetic field dependence of the $D_{5/2}$ Zeeman states
is calculated by diagonalizing the Breit-Rabi Hamiltonian. It
turns out that several transitions starting from one of the
stretched states $S_{1/2}(F=4,m_F=\pm 4)$ become independent of
the first-order Zeeman shift at field values of a few Gauss.
Transitions with vanishing differential Zeeman shifts at non-zero
fields have been investigated in experiments with cold atomic
gases \cite{Harber02} to achieve long coherence times and with
trapped ions \cite{Langer05} for the purpose of quantum
information processing. These transitions are also potentially
interesting for building an optical frequency standard and have
several advantages over $m_F=0\leftrightarrow m_{F^\prime}=0$
transitions. We experimentally confirm our calculations by mapping
the field-dependence of one of these transitions.

%\subsection{\label{sec:level2}Second level heading}
%\subsubsection{\label{sec:level3}Third-level heading: References and Footnotes}

\section{Experimental setup}

%Trap

Our experiments are performed with a single $^{43}$Ca$^+$ ion
confined in a linear Paul trap consisting of two tips and four
blade-shaped electrodes \cite{Schmidt-Kaler03}. A radio frequency
voltage ($\nu_{rf} = 25.642$ MHz; $P_{rf}=7$~W) is fed to a
helical resonator and the up-converted signal is applied to one
pair of blade electrodes while the other blade pair is held at
ground. In such a way, a two-dimensional electric quadrupole field
is generated which provides radial confinement for a charged
particle if the radio frequency and amplitude are chosen properly.
Two stainless steel tips are placed $5$~mm apart in the trap's
symmetry axis and are held at a positive voltage $U_{tips}=1000$ V
providing axial confinement. The electrodes are electrically
isolated by Macor ceramic spacers which assure a 20$\,\mu$m
tolerance in the positioning of the four blades and the tip
electrodes. For the parameters given above, the ion trap confines
a $^{43}$Ca$^+$ ion in a harmonic potential with oscillation
frequencies $\nu_{axial} =1.2$ MHz and $\nu_{radial}=4.2$ MHz in
the axial and radial directions. Micromotion due to stray electric
fields is compensated by applying voltages to two compensation
electrodes. The correct compensation voltages are found by
minimizing the Rabi frequency of the first micromotional sideband
of the quadrupole transition for two different laser beam
directions. The trap is housed in a vacuum chamber with a pressure
of about $10^{-10}$ mbar.

%PI

Single \Ca ions are loaded from an isotope-enriched source (Oak
Ridge National Laboratory; 81.1 \% $^{43}$Ca$^+$, 12.8 \%
$^{40}$Ca$^+$, 5.4 \% $^{44}$Ca$^{+}$) into the trap by
isotope-selective two-step photoionization \cite{Lucas04,
Gulde01}. The first transition from the 4s$\,^1\!$S$_0$ ground
state to the 4p$\,^1\!$P$_1$ excited state in neutral calcium is
driven by an external cavity diode laser in Littrow configuration
at 423~nm. Its frequency is monitored by saturation spectroscopy
on a calcium vapor cell held at a temperature of $300\,^\circ$C
and by a wavelength meter with a relative accuracy of 10 MHz. The
second excitation step connecting the 4p$\,^1\!$P$_1$ state to
continuum states requires light with a wavelength below 390~nm. In
our experiment, it is driven by a free-running laser diode at
375~nm.

%Doppler-cooling and detection

\begin{figure}
\includegraphics[width=8cm]{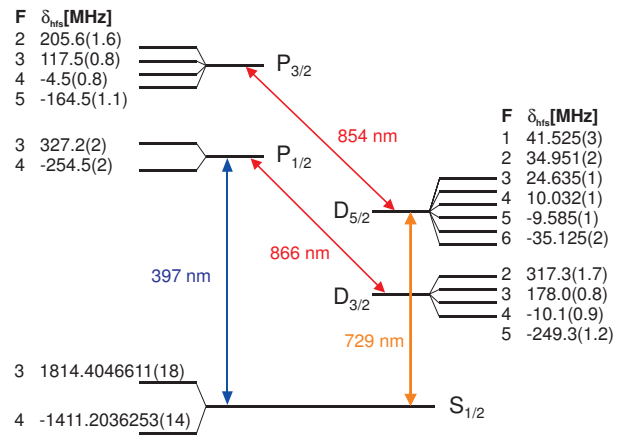}
\caption{\label{fig:levelscheme} (Color online) \Ca level scheme showing the
hyperfine splitting of the lowest energy levels. Hyperfine shifts
$\delta_{hfs}$ of the levels are quoted in MHz (the splittings are
taken from \cite{Werth94,Noertershaeuser98} and our own
measurement). Laser light at 397 nm is used for Doppler cooling
and detection, the lasers at 866 and 854 nm pump out the D-states.
An ultrastable laser at 729 nm is used for spectroscopy on the
quadrupole transition.}
\end{figure}

For laser cooling, a grating-stabilized diode laser is
frequency-doubled to produce light at 397~nm for exciting the
$S_{1/2}\leftrightarrow P_{1/2}$ transition (see Fig.
\ref{fig:levelscheme}). By means of polarization optics and an
electro-optical modulator operated at
$3.2$ GHz, laser beams exciting the following transitions are provided:\\

\begin{tabular}{ccc}
  beam n$^\circ$       & polarization            & transition \\
  \hline
  1, 2\hspace{3pt} & $\pi$, $\sigma^+$       & \hspace{3pt}$\1(F=4) \leftrightarrow \2(F^\prime=4)$ \\
  3 \hspace{3pt} & $\sigma^+$  & \hspace{3pt}$\1(F=3) \leftrightarrow \2(F^\prime=4)$ \\
 \\
\end{tabular}

Laser beams n$^\circ\,1$-3 are all switched on for Doppler cooling
and fluorescence detection. We avoid coherent population trapping
by lifting the degeneracy of the Zeeman sub-levels with a magnetic
field. To avoid optical pumping into the \3 manifold repumping
laser light at 866 nm has to be applied. The repumping efficiency
was improved by tuning the laser close to the
$D_{3/2}(F=3)\leftrightarrow P_{1/2}(F^\prime=3)$ transition
frequency and red-shifting part of the light by $-f_1$,
% $-f_2$,
$-f_1-f_2$ with two acousto-optical modulators (AOMs) operating at
frequencies $f_1=150$~MHz and $f_2=245$~MHz. In this manner, all
hyperfine $D_{3/2}$ levels are resonantly coupled to one of the
$P_{1/2}(F^\prime=3,4)$ levels. Since the electronic g-factor of
the $D_{3/2}(F=3)$ level vanishes, coherent population trapping in
this level needs to be avoided by either polarization-modulating
the laser beam or by coupling the level to both
$P_{1/2}(F^\prime=3,4)$ levels. In our experiment, non-resonant
light ($\delta\approx 190$~MHz) exciting the
$D_{3/2}(F=3)\leftrightarrow P_{1/2}(F^\prime=4)$ seems to be
sufficient for preventing coherences from building up. After
switching off laser beam n$^\circ\,1$, the ion is optically pumped
into the state $S_{1/2}(F=4,m_F=4)$. The pumping efficiency is
better than $95$ \%.

All diode lasers are stabilized to Fabry-Perot cavities. The
cavity spacer is a block of Zerodur suspended in a temperature
stabilized vacuum housing. For frequency tuning, one of the
reference cavity mirrors is mounted using two concentric piezo
transducers that are compensated for thermal drift. This allows
frequency tuning of the lasers over several GHz while achieving low
drift rates (typically $< 100$ Hz/s) once the piezos have settled.

%spectroscopy
To set the magnitude and orientation of the magnetic field, a
single $^{40}$Ca$^+$ ion was loaded into the trap. The ambient
magnetic B-field was nulled by applying currents to magnetic field
compensation coils so as to minimize the ion's fluorescence. After
that, the magnetic field can be set to the desired value by
sending a current through a pair of coils defining the
quantization axis. All coils are powered by homemade current drivers
having a relative drift of less than $2\cdot 10^{-5}$ in $24$~h.

Light for the spectroscopy on the $S_{1/2}\leftrightarrow D_{5/2}$
quadrupole transition is generated by a Ti:Sapphire laser stabilized to
an ultra-stable high finesse reference cavity (finesse
$\mathcal{F}=410000$)\cite{Notcutt05}. The free spectral range of
the cavity was measured to be $\Delta_{FSR}=1933.07309(20)$ MHz by
using a second independently stabilized laser and observing the
beat note for the Ti:Sa laser locked to several different modes.
From this measurement, also an upper limit of less than 50 Hz
could be determined for the laser line width. The frequency drift
of the $729$ nm laser stabilized to the reference cavity is
typically less than $0.5$ Hz/s. By locking the laser to different
modes of the reference cavity and by changing its frequency with
AOMs we are able to tune the laser frequency in resonance with any
transition between levels of $S_{1/2}$ and $D_{5/2}$ in
$^{40}$Ca$^+$ and $^{43}$Ca$^+$. The radio frequencies applied to
the AOMs are generated by a versatile frequency source based on
direct digital synthesis.

Spectroscopy on the quadrupole transition is implemented using a
pulsed scheme. In a first step, the ion is Doppler cooled and
prepared in the $S_{1/2}(F=4,m_F=\pm 4)$ level by optical pumping.
Then the ion is probed on the quadrupole transition by light at
729 nm. At the end of the experimental cycle, the ion's quantum
state is detected by a quantum jump technique. For this, the
cooling laser and the repumper at 866~nm are turned back on for a
duration of 5~ms, projecting the ion onto either the fluorescing
$S_{1/2}$ or the dark $D_{5/2}$ state. The light emitted by the
ion is collected with a customized lens system ($NA=0.27$,
transmission $>~95~\%$) and observed on a photomultiplier tube and
a CCD camera simultaneously. A threshold set for the number of
photomultiplier counts discriminates between the two possibilities
%ion being in state \1 or \5 state
with high efficiency. Finally the \5 state population is pumped
back to \1 by means of another grating-stabilized diode laser
operating at 854 nm. This measurement cycle is repeated a hundred
times before setting the probe laser to a different frequency and
repeating the experiments all over again.
%The recorded spectra look similar to the one shown in
%figure \ref{fig:narrowline}.

%setting the B-field
In order to set the magnetic field precisely, we use a single
$^{40}$Ca$^+$ ion to determine the field strength by measuring the
frequency splitting of the two transitions
$S_{1/2}(m=+1/2)\leftrightarrow D_{5/2}(m^\prime=+5/2)$ and
$S_{1/2}(m=+1/2)\leftrightarrow D_{5/2}(m^\prime=-3/2)$.
%...that the resulting field was $3.3994$ G.
Stray magnetic fields oscillating at multiples of 50~Hz change the
magnitude of the field by less than 2~mG over one period of the
power line frequency. By synchronizing the experiments with the
phase of the power line, ac-field fluctuations at multiples of
50~Hz are eliminated as a source of decoherence. As the duration
of a single experiment typically is on the order of 20~ms, this
procedure does not significantly slow down the repetition rate of
the experiments.

\begin{figure}[t]
\includegraphics[width=8cm]{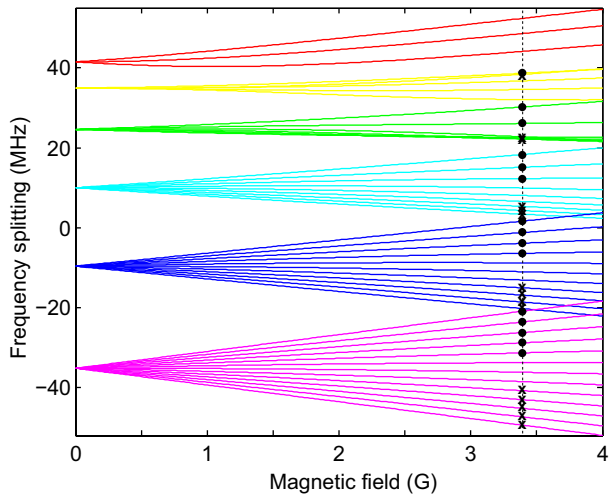}
\caption{\label{fig:D52Manifold} (Color online) Hyperfine and Zeeman splitting of
the $D_{5/2}$ state manifold calculated for hyperfine constants
measured in our experiment. Filled circles (\textbullet) and
crosses ({\bf x}) mark states that can be excited starting from
the $S_{1/2}(F=4)$ state with magnetic quantum number $m_F=+4$
($m_F=-4$), respectively. The vertical dashed line indicates the
magnetic field used for measuring the frequency shifts in the
experiment. \label{Fig2:HyperfineZeeman}}
\end{figure}

\section{Results}

\subsection{Hyperfine coefficients for the \5 state}

The hyperfine structure splitting of the $S_{1/2}$ and $D_{5/2}$
states is determined by effective Hamiltonians \cite{Armstrong71}
$H_{hfs}^{(S_{1/2})}=hA_{S_{1/2}}\mathbf{I\cdot J}$ and, assuming
that $J$ is a good quantum number,
%\begin{equation}
%H_{hfs}^{(D_{5/2})}= hA_{D_{5/2}} \mathbf{I\cdot J}+
%hB_{D_{5/2}}\frac{3(\mathbf{I\cdot
%J})^2+\frac{3}{2}(\mathbf{I\cdot
%J})-I(I+1)J(J+1)}{2I(2I-1)J(2J-1)} \label{HamHFS}
%\end{equation}
\begin{eqnarray}
\label{HamHFS}
H_{hfs}^{(D_{5/2})}&=& hA_{D_{5/2}} \mathbf{I\cdot J}+\\
&+& hB_{D_{5/2}}\frac{3(\mathbf{I\cdot
J})^2+\frac{3}{2}(\mathbf{I\cdot
J})-I(I+1)J(J+1)}{2I(2I-1)J(2J-1)} \nonumber
\end{eqnarray}
operating on the hyperfine level manifolds of the ground and
metastable state. Here $h$ is Planck's constant and
$A_{D_{5/2}}(A_{S_{1/2}})$ and $B_{D_{5/2}}$ are the hyperfine
constants describing the magnetic dipole and electric quadrupole
interactions in the $D_{5/2}$ ($S_{1/2}$) state; higher-order
multipoles \cite{Schwartz55} are not taken into account. Terms
arising from second-order perturbation theory \cite{Schwartz55}
are expected to shift the levels by only negligible amounts as
$(A_{D_{5/2}}\!\cdot\!A_{D_{3/2}})/\Delta_{FS}\approx 100$~Hz
where $\Delta_{FS}$ denotes the fine-structure splitting of the
$D$ states.

In a non-zero magnetic field, the Hamiltonian (\ref{HamHFS}) is
replaced by
\begin{equation}
H^{(D_{5/2})}=H_{hfs}^{(D_{5/2})} + g_{D_{5/2}}\mu_B\mathbf{J\cdot
B} + g_I\prime\mu_B \mathbf{I\cdot B},
\end{equation}
where $g_{D_{5/2}}$ is the electronic g-factor of the $D_{5/2}$
state and $g_I\prime$ denotes the nuclear g-factor. Fig.
\ref{Fig2:HyperfineZeeman} shows the resulting energy shifts of
the Zeeman level caused by hyperfine and Zeeman interactions. The
energies of the $S_{1/2}(F=4,m_F=\pm 4)$ levels used in our
spectroscopic measurements are linearly shifted by
%$h\delta_\pm=\pm(\frac{1}{2}g_{S_{1/2}}\!+\!\frac{7}{2}g_I^\prime)\mu_B
%B$.
$h\delta_\pm=\pm(g_{S_{1/2}}\frac{1}{2}+g_I^\prime I)\mu_B B$.

%30 measurement

From earlier measurements and calculations of the isotope shift
\cite{Werth95} and the hyperfine splitting of the $S_{1/2}$
\cite{Werth94} and the $D_{5/2}$
\cite{Noertershaeuser98,Sahoo03,Yu04} states, the transition
frequencies on the quadrupole transition in $^{43}$Ca$^+$ are
known to within 20~MHz with respect to the transition in
$^{40}$Ca$^+$. This enabled us to unambiguously identify the lines
observed in spectra of the $S_{1/2}\leftrightarrow D_{5/2}$
transition. In a first series of measurements the ion was prepared
in the state $S_{1/2}(F=4,m_F=+4)$ by optical pumping with
$\sigma_+$-polarized light. There are fifteen transitions to the
$D_{5/2}$ levels allowed by the selection rules for quadrupole
transitions. Spectra were recorded on all of them with an
excitation time of 500 $\mu$s in a magnetic field of about 3.40~G.
In a second measurement series, after pumping the ion into
$S_{1/2}(F=4,m_F=-4)$ another fifteen transitions were measured.
To obtain the hyperfine constants of the $D_{5/2}$ state, we
fitted the set of 30 transition frequencies by diagonalizing the
Hamiltonian taking the hyperfine constants $A_{D_{5/2}}$,
$B_{D_{5/2}}$, the magnetic field $B$ and a frequency offset as
free parameters. The hyperfine constant
$A_{S_{1/2}}=-806.4020716$~MHz was measured in reference
\cite{Werth94}. The g-factors $g_I\prime=2.0503\cdot 10^{-4}$
%gI=2.0503%
and $g_{S_{1/2}}=2.00225664$
%gS=2.00225664$
were taken from references \cite{Stone05} and \cite{Tommaseo03},
$g_{D_{5/2}}=1.2003(1)$ was measured by us in an experiment with a
single $^{40}$Ca$^+$ ion. The fit yields
\begin{eqnarray*}
A_{D_{5/2}}&=&-3.8931(2)\textnormal{ MHz,}\\
B_{D_{5/2}}&=&-4.241(4)\,\textnormal{ MHz,}
\end{eqnarray*}
where the standard uncertainty of the determination is added in
parentheses. The average deviation between the measured and the
fitted frequencies is about 1~kHz. If $g_{D_{5/2}}$ is used as a
free parameter, we obtain $g_{D_{5/2}}=1.2002(2)$ and the fitted
values of the hyperfine constants do not change. Also, adding a
magnetic octupole interaction \cite{Schwartz55} to the hyperfine
Hamiltonian does not change the fit values of the hyperfine
constants.

%isotope shift
\subsection{Isotope shift}
After having determined the values of $A_{D_{5/2}}$ and
$B_{D_{5/2}}$, the line center of the \Ca \1 $\leftrightarrow$~\5
transition can be found. By comparing the transition frequencies
in $^{43}$Ca$^+$ and in $^{40}$Ca$^+$, the isotope shift
$\Delta_{iso}^{43,40}=\nu_{43}-\nu_{40}$ is determined. Switching
the laser from $\nu_{40}$ to $\nu_{43}$ is achieved by locking the
laser to a $TEM_{00}$ cavity mode three modes higher
($\nu_{n+3}=\nu_n+3\Delta_{FSR}$) than for $^{40}$Ca$^+$
 and adjusting its frequency with an AOM. For
the isotope shift, we obtain
\begin{equation*}
\Delta_{iso}^{43,40}=4134.713(5)\textnormal{ MHz.}
\end{equation*}
This value is in good agreement with a previous measurement
($\Delta_{iso}^{43,40}=4129(18)$~MHz \cite{Werth94}). Frequency
drift between the measurements, accuracy of the reference cavity's
free spectral range $\Delta_{FSR}$ and the uncertainty in the
determination of the exact line centers limit the accuracy of our
measurement.

%b-insensitive line
\subsection{Magnetic field independent transitions}
\begin{figure}[t]
\includegraphics[width=8cm]{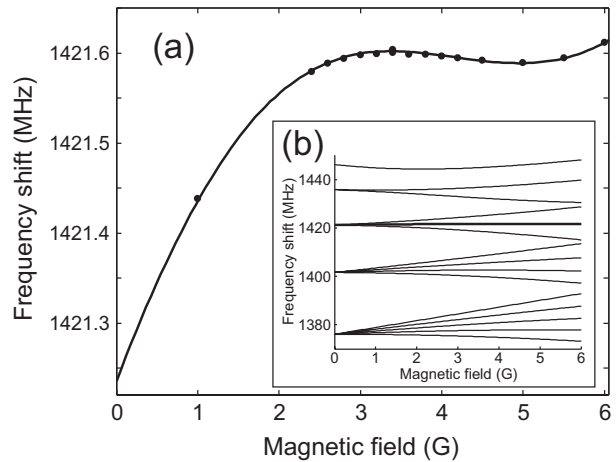}
\caption{\label{fig:Fieldindependent} (a) Frequency dependence of
the $S_{1/2}(F=4,m_F=4) \leftrightarrow D_{5/2}(F=4,m_F=3)$
transition frequency for low magnetic fields. The transition
frequency becomes field-independent at B=3.38~G and B=4.96~G with
a second-order Zeeman shift of $\pm 16\textnormal{kHz/G}^2$. The
measured data are not corrected for the drift of the reference
cavity which may lead to errors in the shift of about 1-2 kHz. To
match the data with the theoretical curve based on the previously
measured values of $A_{D_{5/2}}$, $B_{D_{5/2}}$, an overall
frequency offset was adjusted. (b) Calculated shift of the fifteen
allowed transitions starting from $S_{1/2}(F=4,m_F=4)$. The thick
line shows the transition to the state
$D_{5/2}(F^\prime=4,m_{F^\prime}=3)$.}
\end{figure}

Given the measured values of the hyperfine coefficients
$A_{D_{5/2}}$, $B_{D_{5/2}}$, we calculate that there are seven
transitions starting from the stretched states $S_{1/2}
(F=4,m_F=\pm 4)$ that have no first order Zeeman effect for
suitably chosen magnetic fields in the range of 0-6~G. These
transitions are useful as they offer the possibility of measuring
the line width of the spectroscopy laser in the presence of
magnetic field noise. To demonstrate this property, we chose the
transition $S_{1/2}(F=4,m_F=4) \leftrightarrow D_{5/2}(F=4,m_F=3)$
which has the lowest second-order dependence on changes in the
magnetic field. We measured the change in transition frequency for
magnetic fields ranging from one to six Gauss as shown in figure
\ref{fig:Fieldindependent}. The black curve is a theoretical
calculation based on the measurement of the hyperfine constants.
For the data, the frequency offset is the only parameter that was
adjusted to match the calculated curve. Both, the experimental
data and the model show that the transition frequency changes by
less than 400~kHz when the field is varied from 0 to 6 G. The
transition frequency becomes field-independent at about B=3.38 G
with a second order B-field dependency of -16 kHz/G$^2$, which is
six times less than the smallest coefficient for a clock
transition based on $m_F=0\leftrightarrow m_{F^\prime}=0$
transitions at zero field. At B=4.96~G the linear Zeeman shift
vanishes again.

\begin{figure}[t]
\includegraphics[width=8cm]{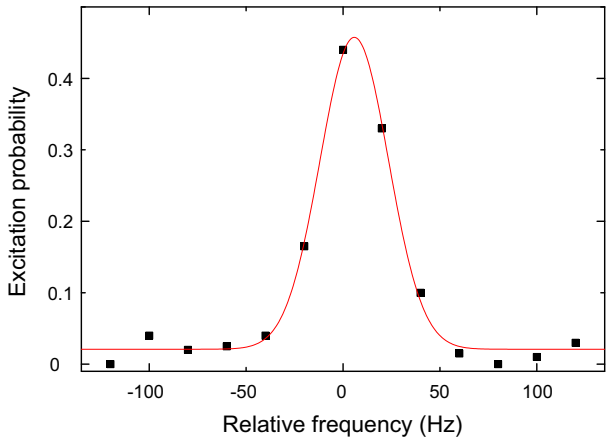}
\caption{\label{fig:narrowline} (Color online) Frequency scan over the transition
$S_{1/2}(F=4,m_F=4) \leftrightarrow
D_{5/2}(F^\prime=4,m_{F^\prime}=3)$ with an interrogation time of
100~ms. A Gaussian fit (solid line) determines a width of 42~Hz
which is dominated by the line width of the spectroscopy laser at
729~nm.}
\end{figure}

We used the field-independence of this transition for
investigating the phase coherence of our spectroscopy laser. For
this, we set the magnetic field to 3.39 G and recorded an
excitation spectrum of the transition by scanning the laser over
the line with an interrogation time of $100$ ms. The result is
depicted in figure \ref{fig:narrowline}. A Gaussian fit gives a
line width of 42 Hz. The observed line width is not yet limited by
the life time $\tau$ of the \5 state ($\tau = 1.17 s)$ or by the
chosen interrogation time. Line broadening caused by magnetic
field fluctuations can be excluded on this transition. Also,
ac-Stark shifts are expected to play only a minor role. Therefore,
we believe that the observed line width is mostly related to the
line width of the exciting laser.

%\section{Discussion}

\section{Summary and discussion}
The hyperfine structure of the $D_{5/2}$ level in $^{43}$Ca$^+$
has been observed and precisely measured by observing frequency
intervals of the $S_{1/2}(F=4,m_F=\pm 4)\leftrightarrow
D_{5/2}(F'=2\ldots 6,m_{F^\prime})$ transitions at non-zero field.
These measurements yielded values for the hyperfine constants
$A_{D_{5/2}}$, $B_{D_{5/2}}$ as well as a determination of the
isotope shift of the quadrupole transition with respect to
$^{40}$Ca$^+$. A diagonalization of the $D_{5/2}$ state's
Hamiltonian showed that several transitions exist which become
magnetic-field independent at small but non-zero values of $B$.
These transitions are of practical importance for probing the
laser line width of the spectroscopy laser and for monitoring the
drift rate of its reference cavity. For the purpose of building an
optical frequency standard based on $^{43}$Ca$^+$
\cite{Champenois04, Hosokawa05}, they might be superior to the
transitions $S_{1/2}(F,m_F=0)\leftrightarrow D_{5/2}(F^\prime,
m_{F^\prime}=0)$ for the following reasons: (i) the initialization
step requires only optical pumping to the stretched state
$S_{1/2}(F=4,m_F=\pm 4)$ which can be conveniently combined with
resolved sideband cooling to the motional ground state of the
external potential. (ii) The magnetic field can be exactly set to
the value where the transition becomes field-independent while
still maintaining a well-defined quantization axis. (iii) The
second-order Zeeman effect can be reduced to a value that is six
times smaller than what can be achieved for the best
$m_F=0\leftrightarrow m_{F^\prime}=0$ 'clock transition'. Still,
we are somewhat cautious about the usefulness of $^{43}$Ca$^+$ as
an optical frequency standard as compared to other candidate ions.
While the rather small hyperfine splitting of the metastable state
has the nice property of providing field-independent transitions
at low magnetic fields, it risks also to be troublesome as the
induced level splitting is about the same size as typical trap
drive frequencies. Improperly balanced oscillating currents in the
trap electrodes might give rise to rather large ac-magnetic level
shifts.

\begin{acknowledgments}
We wish to acknowledge support by the Institut f\"ur
Quanteninformation GmbH and by the U. S. Army Research Office. We
acknowledge P.~Pham's contribution to the development of a
versatile source of shaped RF pulses.
\end{acknowledgments}

\bibliography{apssamp}% Produces the bibliography via BibTeX.

\end{document}